# On some pitfalls of the log-linear modeling framework for capture-recapture studies in disease surveillance

Yuzi Zhang[1,*], Lin Ge[1], Lance A. Waller[1], and Robert H. Lyles[1]
[1]Department of Biostatistics and Bioinformatics, The Rollins School of Public Health of Emory University, 1518 Clifton Rd. N.E., Atlanta, GA, USA
*yuzi.zhang@emory.edu

## Abstract

In epidemiological studies, the capture-recapture (CRC) method is a powerful tool that can be used to estimate the number of diseased cases or potentially disease prevalence based on data from overlapping surveillance systems. Estimators derived from log-linear models are widely applied by epidemiologists when analyzing CRC data. The popularity of the log-linear model framework is largely associated with its accessibility and the fact that interaction terms can allow for certain types of dependency among data streams. In this work, we shed new light on significant pitfalls associated with the log-linear model framework in the context of CRC using real data examples and simulation studies. First, we demonstrate that the log-linear model paradigm is highly exclusionary. That is, it can exclude, by design, many possible estimates that are potentially consistent with the observed data. Second, we clarify the ways in which regularly used model selection metrics (e.g., information criteria) are fundamentally deceiving in the effort to select a "best" model in this setting. By focusing attention on these important cautionary points and on the fundamental untestable dependency assumption made when fitting a log-linear model to CRC data, we hope to improve the quality of and transparency associated with subsequent surveillance-based CRC estimates of case counts.

## Introduction

Capture-recapture (CRC) methods are widely applied when analyzing case identification data collected from multiple overlapping surveillance systems to estimate the number of prevalent or incident cases ($N$) across a study period among essentially closed populations. When applying statistical models, at least one crucial but untestable assumption is necessary for identifying $N$, except under very specific circumstances (Lyles et al. 2021; Chao, Pan, and Chiang 2008). To implement the CRC method, surveillance data are generally summarized in a frequency table based on the possible capture histories across surveillance systems. The key challenge in CRC follows from the fact that this frequency table is incomplete since the number of cases not captured by any system is missing.

For CRC methods applied to estimate disease risk or prevalence in human populations, log-linear models are widely adopted assuming independent Poisson distributions for the observed cell counts (Fienberg 1972; Cormack 1989). To estimate the total number of cases, the log-linear model regresses observed cell counts against indicators of the capture history and their interactions. The popularity of log-linear models is partially due to their allowance for dependency between systems via interaction terms. In addition, the accessibility of log-linear models also greatly promotes their applications. For example, the `GENMOD` procedure in SAS and the `glm` function in R can be directly used for fitting log-linear models to CRC data (SAS 2013; R Development Core Team 2021).

The application of log-linear models has been extended to incorporate covariates. Discrete covariates can be incorporated by stratification of the cell counts, or by directly including the covariates and their interactions with indicators of the capture history in the log-linear model (Fienberg 1972; Hook and Regal 1995). The log-linear model is a discrete cell-count model, and hence can only incorporate continuous covariates by first categorizing them. However, logistic models allow one to include both individual-level discrete and continuous covariates to model the capture probability of each data stream by assuming independence conditional on covariates (Huggins 1989; Alho 1990). Further extensions of this model permit consideration of the dependency structure between data streams while also modeling the individual-level capture probabilities using the logistic model (Zwane and van der Heijden 2005). In this model, the dependency between data streams is induced by allowing capture probabilities to depend on covariates and responses to data streams.

In epidemiological studies, practitioners usually undertake the fitting of all possible log-linear models after excluding the highest-order interaction term. This exclusion is by common convention as a means of ensuring identifiability, since the number of cases not captured by any system is unobserved. Subsequently, the estimated $N$ and associated 95% confidence interval (CI) are typically reported based on the model identified by a model selection metric, e.g., Akaike's information criterion (AIC), likelihood ratio test statistics, or the Bayesian Information Criterion (BIC) (Akaike 1974; Héraud-Bousquet et al. 2012; Hook and Regal 1997; Schwarz 1978; Barocas et al. 2018). Generally, the log-linear model with the lowest AIC will be selected as the "best" model to obtain the estimate of $N$ (Hook and Regal 1997; Héraud-Bousquet et al. 2012). However, it has been noted in various applied studies that different log-linear models can result in widely different estimated $N$ (Poorolajal, Mohammadi, and Farzinara 2017; Zhang and Small 2020; Ramos et al. 2020). In other words, the estimation of $N$ can be very sensitive to model assumptions that cannot be verified using only the observed data (Fienberg 1972; Hook and Regal 1997; Lyles et al. 2021). However, those commonly applying log-linear models for CRC-based estimation in current practice are often unaware of the extent to which regularly-used model selection metrics based on the observed data are inadequate.

In this paper, we illustrate two major pitfalls associated with the log-linear model paradigm for CRC, primarily stemming from the fact that it dissuades users from carefully considering or attempting to vet the assumptions (e.g., regarding population-level dependencies among data streams) required to enable the estimation of $N$. First, we show that the log-linear model framework is highly “exclusionary”. Specifically, the log-linear model generally excludes many possible estimates of $N$ by design (as illustrated below). Second, we illustrate how model selection metrics can be deceptive and inadequate tools for CRC model selection. We clarify how such metrics choose models based on certain observable constraints in the data, despite the fact that these constraints actually provide no information about true non-identifiable dependency parameters characterizing the data streams that are central to the valid estimation of $N$. Our goal is to caution that application of the log-linear model framework to analyze CRC data requires a clear understanding of these pitfalls, with careful consideration of what key dependency assumptions are imposed (sometimes implicitly) by the adopted model.

## Motivating Data

We used two publicly available epidemiological CRC datasets for quantifying HIV infections, together with one dataset simulated in a setting where referrals between data streams occur (Jones et al. 2014). The data demonstrate how the log-linear model framework excludes many possible estimates of $N$ by design. The two real data examples involve three and four data streams, respectively, and both were analyzed previously using the standard log-linear model paradigm (Abeni, Brancato, and Perucci 1994; Poorolajal, Mohammadi, and Farzinara 2017). Specifically, the three-stream HIV CRC data were collected in Iran in 2016 and the four-stream HIV data were collected in Lazio, Italy during 1990.

## MLEs of $N$ with a Given Key Dependency Parameter

We first review the maximum likelihood estimator (MLE) based on a population-level multinomial model leveraging conditional inference under the two-stream case, and then extend to incorporate multiple (>2) data streams (Darroch 1958). Prior work has demonstrated that the population-level multinomial model can be generated from the individual-level multinomial model, where the vector of indicators for capture histories of an individual is assumed to follow a multinomial distribution with size 1 (Lyles et al. 2021). We assume a closed population with $N$ diseased cases, with $K$ surveillance streams implemented. Under this situation, a total of $2^K$ unique capture histories exist. Let the set $\mathcal{O} = \{(1,1,\dots,1),\dots,(0,0,\dots,0)\}$ containing $2^K$ sequences denote the collection of all possible capture histories, where capture histories are arranged in lexicographic order and 1 denotes captured and 0 not captured. Let $h_j$ denote the $j$-th sequence in the set $\mathcal{O}$ with $j = 1,\dots,2^K$, $N_{h_j}$ and $n_{h_j}$ denote the true and observed number of cases having capture history $h_j$. Because the number of cases not captured by any stream is unobserved, $n_{(0,0,\dots,0)}$ is missing. Hereafter, we remove parentheses and commas from subscripts.

When $K = 2$, one can introduce a parameter vector $(p_1, p_{2|1}, p_{2|\bar{1}})$ to model the three observed cell counts $(n_{11}, n_{10}, n_{01})$ under the population-level multinomial model conditional on being captured while allowing an arbitrary level of dependency between two streams (Lyles et al. 2021). For a given value of $\psi = p_{2|\bar{1}}$, the MLE of $N$ is given by

$$\widehat{N}_\psi = n_{11} + n_{10} + \frac{n_{01}}{\psi}, \tag{1}$$

where $\psi$ is interpreted as the population-level probability of capture by stream 2 given a lack of capture by stream 1. Note that $\psi$ is inestimable based solely on the observed data. The dependency between two streams can also be quantified by the ratio $\phi = p_{2|1}/p_{2|\bar{1}}$, where $\phi = 1$ implies the two streams operate independently at the population level, $> 1$ implies a positive dependency, and $< 1$ implies a negative dependency. For a known $\phi$, the MLE of $N$ is (Chen 2020; Lyles et al. 2021):

$$\widehat{N}_{\phi} = n_{11} + n_{10} + \frac{n_{01}(n_{11} + n_{10})}{n_{11}}\phi. \tag{2}$$

It is worth noting that *any* valid value of $\psi$ or $\phi$ yields exactly the same maximized log-likelihood value. Thus, the MLEs in Equations (1) and (2) are equally consistent with the observed data, as long as one supplies a feasible value of $\psi$ or $\phi$.

For the case of $K$ streams, the extended parameter vector $(p_1, p_{2|1}, p_{2|\bar{1}}, p_{3|12}, p_{3|1\bar{2}}, p_{3|\bar{1}2}, \dots, p_{K|\overline{12\dots,K-1}})$ models the observed $2^K - 1$ cell counts, where $p_1$ denotes the marginal probability of identification by the first stream and the remaining parameters are population-level conditional probabilities of identification in the $i$-th stream given all other capture histories arranged in lexicographic order ($\bar{\cdot}$ indicates no, otherwise yes). For example, $p_{3|1\bar{2}}$ represents the proportion captured by the third stream given captured by the first stream but not captured by the second. As noted above, in the two-stream case we refer to $p_{2|\bar{1}}$ as the key inestimable parameter and use $\psi$ to denote it. Similarly, we denote $p_{K|\overline{12\dots,K-1}}$ as $\psi$ and treat it as the key inestimable parameter when $K > 2$.

Given $K > 2$ streams, we derive the MLE of $N$ for a known value of $\psi$ and its variance estimator under the population-level conditional multinomial model as follows:

$$\widehat{N}_{\psi} = \sum_{\{j:h_j \in \mathcal{O}^*\}} n_{h_j} + \frac{n_{00\dots1}}{\psi}, \tag{3}$$

$$\widehat{Var}(\widehat{N}_{\psi}) = \frac{1-\psi}{\psi^2} n_{00\dots1}, \tag{4}$$

where $\mathcal{O}^* = \mathcal{O} \setminus \{(0,0,\dots,0),(0,0,\dots,1)\}$, and $\sum_{\{j:h_j \in \mathcal{O}^*\}} n_{h_j}$ is the observed number of cases caught at least once minus the observed number of cases caught by the last stream but not by any other stream. This MLE and its variance estimator under the three-stream case have been proposed previously (Zhang et al. 2023), but Equations (3)-(4) extend the result to incorporate an arbitrary number of streams. As in the two-stream case, we emphasize that $\widehat{N}_{\psi}$ is equally consistent with the observed cell counts for any valid value of $\psi$ (i.e., $0 < \psi \leq 1$) that one supplies. That is, the parameter $\psi$ is non-identifiable based on the observed data alone.

## The Exclusionary Property of CRC Log-linear Models

The log-linear model assumes that cell counts are independent Poisson variables conditional on model coefficients that determine their corresponding means. For the simplest CRC setting with two-stream data and no covariates, the corresponding fully specified log-linear model is:

$$\log\left[E\left(N_{h_j}\right)\right] = \alpha + \beta_1 X_1 + \beta_2 X_2 + \delta X_1 X_2, \tag{5}$$

where $X_k = 1$ if cases are identified by the $k$-th stream under capture history $h_j$, and 0 otherwise, for $k = 1,2$. The estimated $N$ is then computed as $n_c + \exp(\hat{\alpha})$, where $n_c$ is the number of cases captured at least once and $\hat{\alpha}$ is the MLE of the intercept. With 3 observed cell counts, it is impossible to estimate all 4 model coefficients. In the above example, the standard CRC log-linear modeling convention of dropping the $K$-way interaction term implies setting $\delta = 0$. In addition, a hierarchy principle is also typically applied, whereby any log-linear model considered must include all possible lower level interaction terms before the inclusion of any higher level interaction term. When $K > 2$, a valid candidate model also must include all first-order terms. Applying the first convention for the basic two-stream case, only 3 models would therefore be valid candidates; however, note that there are a total of 7 possible models that can be fitted if these conventions are not applied. Each of these has been closely examined by prior authors, who showed that the full set of 7 possible models allows for at most 4 unique estimates of the ratio parameter $\phi$ (see Table 2 in Lyles et al., 2021). Yet, as noted earlier, there are in fact an infinite number of values of the key parameter ($\psi$ or $\phi$) that yield unique MLEs of $N$ and are equally consistent with the observed data. This simple special case exhibits the exclusionary nature of CRC log-linear models, in the sense that they exclude many possible estimates by design. As we will see, this important pitfall is not avoided simply by including additional streams and/or covariates.

*A Toy Example*

To illustrate the exclusionary nature of the log-linear model, we apply the novel visualization tool proposed in Zhang et al. (2023) to the toy set of cell counts presented in Table 1. In this example, we deliberately set $n_{11}$ equal to $n_{01}$ in order to demonstrate other pitfalls discussed in later sections.

Figure 1 shows estimates from all 7 possible log-linear models as well as MLEs of $N$ in Equations (1)-(2) with various assumed $\psi$ and $\phi$ values based on the example data. The figure demonstrates that the log-linear model framework excludes large swaths of possible estimates of $N$. In this example, considering all possible log-linear models only results in 4 unique estimates of $N$, despite the fact that a continuum of estimated $N$ values is consistent with the observed data. As shown in Figure 1(B), the 7 possible log-linear models only permit $\phi$ to take estimated values of 2/3, 4/5, and 1. Yet, any value of $\phi$ within the range $[1/3, \infty)$ would result in an equally valid estimate of $N$ based on the observed data. The lower bound of $1/3$ is derived based on the natural lower bound of $N$, i.e., $N$ is no smaller than the number of cases captured at least once.

It is also noteworthy that, in this example, none of the possible log-linear models allows positive dependency between data streams (i.e., $\phi > 1$). However, there is no evidence in the observed data to exclude this possibility, in light of the unobserved cell count $n_{00}$. Moreover, two positively correlated streams are often seen in practice (Hook and Regal 2000). For instance, both streams may tend to identify cases with similar characteristics.

Although it has been shown that MLEs of $N$ derived using the population-level multinomial model are equivalent to estimates yielded by log-linear models when equivalent assumptions are imposed (Cormack and Jupp 1991), we observe that there are estimates from unsaturated (i.e., the number of model coefficients is less than the number of observed capture histories) log-linear models (e.g., M1, M4) that do not appear on the curve continuum in Figure 1. This is because unsaturated log-linear models also impose assumptions on observed cell counts, while the MLEs of $N$ assuming known $\psi$ or $\phi$ do not. That is, when the log-linear model imposes constraints on observed cell counts, the two models indeed impose different sets of assumptions. For example, model 1 constrains the three fitted cell counts to be equal (i.e., $p_{11} = p_{10} = p_{01}$), so that the fitted counts differ from the observed in this example. Since fitted cell counts always coincide with observed cell counts when fitting saturated log-linear models, estimates from those models are consistent with the MLEs in Equations (1) and (2) when fixing $\psi$ or $\phi$ at the corresponding log-linear model-based estimates. In this example where $n_{11} = n_{01}$, the unsaturated model 3 happens to result in an estimated $N$ equal to an MLE based on Equation (2), as shown in Figure 1(B). However, when $n_{11}$ is not equal to $n_{01}$, estimates from model 3 will deviate from the MLE.

Importantly, the log-linear model framework continues to exclude many valid estimates even though the number of possible log-linear models increases exponentially with the number of data streams. For the three-stream case, the fully specified log-linear model is given by

$$\log\left[E\left(N_{h_j}\right)\right] = \alpha + \beta_1 X_1 + \beta_2 X_2 + \beta_3 X_3 + \gamma_1 X_1 X_2 + \gamma_2 X_1 X_3 + \gamma_3 X_2 X_3 + \delta X_1 X_2 X_3. \quad (6)$$

With 7 observed cell counts, a total of 127 log-linear models are possible, of which only 8 are valid candidates when applying both standard conventions discussed previously. While model (6) allows both positive and negative associations between data streams by including pairwise interaction terms, studies have pointed out that the log-linear model framework is incapable of incorporating certain dependency scenarios (Jones et al. 2014).

To expand on this point, Figure 2 displays a variety of estimated $N$ values obtained under different assumptions using log-linear models and the MLE in Equation (3), based on simulated three-stream data analyzed in Jones et al. (2014). These data were simulated from a population-level multinomial distribution by assuming $N = 200{,}000$, the first stream (S1) and the third stream (S3) are independent conditional on the second stream (S2), and that 20% of cases captured by S1 are referred to S3. In Figure 2, while we cut $\psi$ at the values of 0.02 and 0.9, it is possible that $\psi$ can go beyond 0.9 (in which case the estimated $N$ would remain flat) or below 0.02 (in which case the estimated $N$ could theoretically blow up). It is clear that estimates of $N$ from log-linear models only occupy a small part of the curve. For example, no log-linear model places the estimated $N$ within the range 310,000 to 860,000, despite the fact that all values in that range (and all other values depicted on the curve) are equally consistent with the observed data. A key point here (explored further below) is that the log-linear modeling framework leads the user to believe that such exclusions are based on relevant information available in the data, but this is not the case. In addition, as prior authors (Jones et al. 2014) concluded, no log-linear model captures the true associations among the three streams under which these data were generated.

We also explored actual three-stream CRC data analyzed in a previous study (Poorolajal, Mohammadi, and Farzinara 2017). In Figure 3, we observe for example that there is no log-linear model implying the potentially reasonable assumption that $\psi = p_{3|1\bar{2}}p_{3|\bar{1}2}/p_{3|12}$, i.e., the population-level association between S2 and S3 is the same regardless of identification (or not) by S1. Similarly, no model reflects the assumption $\psi = p_{3|12}$, i.e., the proportion captured by S3 among those not captured by S1 and S2 is equal to the proportion captured by S3 among those captured by both S1 and S2. As shown in Table S2, the 8 possible models only yield at most 8 unique estimates of $N$ when applying the usual CRC conventions. Again, however, a continuum of estimates is consistent with the observed data and a significant swath of these estimates are unachievable when fitting all possible log-linear models.

We note that the number of possible log-linear models increases exponentially even when the standard conventions are applied. With four data streams, there are 113 possible log-linear models under the standard conventions (Hook and Regal 1995), while there are a total of 32,767 models if we consider all possible combinations of predictors. However, with this many possible models, the log-linear model framework still excludes many possible estimates as illustrated using data from Abeni, Brancato, and Perucci (1994) (Figure 4). For example, no log-linear model projects $\psi = p_{4|\overline{123}}$ into the range from 0.06 to 0.17. However, this range in fact contains many plausible assumptions, including for example that the key parameter $\psi$ is equal to $p_{4|\bar{1}23}$ or $p_{4|1\bar{2}3}$.

For cases involving more than two streams (Figures 2-4), note again that not all estimates from log-linear models are on the black curve. As discussed in the two-stream case, log-linear models that impose the same assumption about the key parameter $\psi$ do not necessarily impose the same assumptions about observed cell counts due to constraints built into unsaturated models.

## AIC is Deceiving as a Metric for CRC Model Selection

To settle upon a final model for estimating $N$ under the log-linear model framework, practitioners often rely on common metrics such as AIC, BIC, or likelihood ratio test statistics. However, numerous sources have raised concerns about the adequacy of such metrics in CRC modeling for epidemiologic surveillance (Fienberg 1972; Hook and Regal 1997; Lyles et al. 2021; Coull and Agresti 1999).

The main issue associated with typical model selection metrics in CRC settings is that they attempt to identify a “best-fitting” model based solely on the observed data, which in themselves contain no information about any key dependency parameter upon which assuredly valid estimation and inference could be based. That is, certain *untestable* assumptions (e.g., regarding the dependency structure between streams) are necessary to identify an estimate of $N$, but these metrics are incapable of assessing untestable assumptions. Nonetheless, a metric like AIC will happily select a log-linear model on the basis of certain *testable* assumptions. The problem is that such a model then projects an assumption about a non-identifiable dependency parameter that fundamentally determines the estimate of $N$. Unfortunately, the unsuspecting user is typically unaware of the exclusionary property, or the fact that this projection is little more than a mathematical construct. Indeed, the log-linear model that fits the observed data best might project an assumption about dependencies among data streams which differs greatly from the underlying truth. That underlying truth could be much more consistent with the estimated $N$ from a poorly fitting log-linear model, or, as we have seen, might not be attainable by any log-linear model.

Table S1 presents estimates along with the AIC for each model fitted to the toy example data in Table 1. Note that model 3 yields the lowest AIC; this occurs because $n_{11} = n_{01}$ in the observed data. The corresponding mathematical construct, for which there is no basis, then projects the unobserved cell count $n_{00}$ to be the same as the observed cell count $n_{10}$. The key point is that the testable constraint $E(N_{11}) = E(N_{01})$ producing the selected model can be defended based on the data. However, the assumption $E(N_{10}) = E(N_{00})$ is untestable in the observed data. Suppose the true value of $\phi$ were 1.5, which indicates $E(N_{00}) = 875$ and $E(N) = 1{,}875$. This truth differs greatly from the model-projected estimate of $N_{00}$, which equals 500, from applying model 3.

Importantly, the problem associated with relying on testable constraints to project the unobserved cell count remains even when the number of streams increases. Using model 2 in Table S2 as an example, the estimated unobserved cell count $N_{000}$ takes its form only because the testable constraints $E(N_{111})/E(N_{110}) = E(N_{101})/E(N_{100}) = E(N_{011})/E(N_{010})$ are assumed under model 2. The computation of $N_{000}$ is a purely mathematical construct, and the untestable assumption used for inferring $N_{000}$ cannot be justified using the observed data. Yet, a practitioner selecting model 2 based on a metric like AIC would be led to believe otherwise.

For the three-stream CRC data reflected in Figures 2 and 3, Tables 2 and 3 present results from the candidate log-linear models following the standard conventions, together with those models (among all possible) that yield the most favorable AIC. Table 2 illustrates that all saturated models fit the data equivalently, despite the fact that the resulting estimates range from 152,000 to 877,000. No unsaturated model “beats” those saturated models in terms of AIC. When following the standard conventions, model 8 (resulting in an estimate far from the true $N = 200{,}000$) would be selected. We note again that these data were simulated under a referral scenario that no log-linear model is able to incorporate; thus, researchers have no chance of approaching the true dependency structure even if they carefully follow the recommended log-linear model paradigm. Similarly, we observe that model 9 in Table 3 is an unsaturated model which yields lower AIC than the saturated models. This model assumes the testable constraint, $E(N_{110})/E(N_{101}) = E(N_{010})/E(N_{001})$, and projects the unobserved count based on the subsequent assumption $E(N_{000}) = E(N_{100})E(N_{001})/E(N_{101})$. While the observable constraint might be supported based on the observed data, the assumption about the unobserved count is a mathematical construct that could not actually be justified without external knowledge about operating characteristics of the surveillance streams.

We conducted additional simulation studies to further demonstrate that AIC is not a reliable tool for CRC model selection. These simulations focus on the three-stream case, which is a common scenario in epidemiological studies. We simulated 1,000 datasets from a population-level multinomial model under two different scenarios assuming $N = 5{,}000$. In the first scenario, one testable assumption and one untestable assumption were imposed, while in the second scenario, two testable assumptions and one untestable assumption were imposed (details of these simulation settings can be found in Supplementary Materials).

Under scenario 1, it is easy to show that the model that was most frequently ($\sim 80\%$) selected by the AIC implies the testable assumption used for data generation (Table 4). The untestable assumption projected by this model suggests that the key parameter $\psi = p_{3|\overline{12}} = 0.5$. However, the data were generated assuming this key parameter equals $0.1$. Hence, it is not surprising that the average estimate obtained from the selected model (3,040) is far afield from the true value of 5,000. The other two models that were selected by the AIC under some simulations also project very different estimates of $N$. In similar fashion, the two most frequently selected models under scenario 2 permit the two prespecified testable assumptions when certain constraints are applied to model coefficients. Again, the average estimates obtained from the two most frequently selected models (6,848 and 4,407) fall far from the truth because neither model projects the correct untestable assumption.

When applying the standard conventions under scenario 1, note that the saturated model was almost always selected as "best". However, the corresponding averaged estimated $N$ is 8053, greatly overestimating the true value of 5000. All of these simulation results highlight the fact that using AIC as a CRC model selection metric is misleading, since models fitting the observed data "best" cannot in fact be assumed to reliably project the correct untestable assumption which is critical for estimating $N$.

## Discussion

In this work, we first demonstrated that the CRC log-linear model framework is highly exclusionary, in the sense that applying that framework can exclude, by design, broad ranges of estimates of $N$ that are in fact as consistent with the observed data as any others. In practice, we believe any CRC model framework that ignores potentially valid estimates before they can even be considered should be used with great caution. For example, prior authors suggest that epidemiological CRC data streams often tend to reflect a net positive dependence (Hook and Regal 2000). However, as we have seen in Figures 1-4, log-linear models might exclude many feasible estimates corresponding to the situation in which data streams are positively correlated.

As stated in Coull and Agresti (1999), the dependency between data streams cannot reliably be uncovered using only the observed cell counts (Coull and Agresti 1999). We have further emphasized this point via simulation studies demonstrating that the model selection procedure relying on the metric AIC is fundamentally deceiving. Specifically, the model that fits the observed data "best" cannot be assumed to project the correct untestable assumption about the unobserved cell count. Other model selection metrics (e.g., BIC) would suffer from the same problem. It is also worth noting that adding extra data streams cannot necessarily provide information for reliably inferring the dependency among existing data streams. Along these lines, it can be tempting to infer pairwise dependences using three-stream data (Lum and Ball 2015). However, no matter how many data streams are included, the number of cases not captured by any streams is still unobserved. Thus, dependencies among data streams typically remain unclear. Because of the exclusionary property and the fact that model selection in the CRC log-linear model framework is fraught with the pitfalls exhibited herein, we encourage concerted efforts toward a departure from the current reality in which that framework is the centerpiece of standard practice for CRC-based epidemiological surveillance.

As one step toward such a departure, we note that the novel visualization plots based on the MLEs in Equations (1)-(3) have been demonstrated to permit a continuum of estimates of $N$. These MLEs potentially shed light on an alternative modeling framework that would avoid the "exclusionary" problem of the log-linear model. In terms of model selection, an accessible alternative framework focused on leveraging an epidemiologist's knowledge about operating characteristics of surveillance streams would be attractive, since (despite what metrics like AIC might suggest) no information about the true dependencies among streams is available in the observed data alone. A key benefit of such an approach would be the ability to foster careful consideration and transparency with regard to the untestable assumption(s) upon which estimation is based, as opposed to having them dictated as murky mathematical constructs based on the modeling framework that one adopts. Nevertheless, we believe the nature of CRC surveillance will almost always suggest a role for sensitivity and/or uncertainty analysis (Zhang and Small 2020; Zhang et al. 2023).

Table 1: Toy two-stream CRC data

| | Captured by Stream 2 | | |
|---|---|---|---|
| Captured by Stream 1 | Yes | No | |
| Yes | $n_{11} = 250$ | $n_{10} = 500$ | $n_{1.} = 750$ |
| No | $n_{01} = 250$ | $n_{00} = ?$ | |
| | $n_{.1} = 500$ | | $N = ?$ |

* $n_{1.} = n_{11} + n_{10}$ and $n_{.1} = n_{11} + n_{01}$

Table 2: Log-linear models for three-stream data simulated by assuming $N = 200{,}000$ and analyzed by Jones et al. (2014) under the usual conventions and with most favorable AIC

| **Fitted models under the usual conventions** | | | | | | | | | | |
|---|---|---|---|---|---|---|---|---|---|---|
| Model | $X_1$ | $X_2$ | $X_3$ | $X_1X_2$ | $X_1X_3$ | $X_2X_3$ | $X_1X_2X_3$ | $\hat{N}$ | $\hat{\psi}$ | AIC |
| 1 | × | × | × | | | | | 199,910 | 0.446 | 51,619 |
| 2 | × | × | × | × | | | | 260,386 | 0.342 | 18,469 |
| 3 | × | × | × | | × | | | 196,734 | 0.461 | 51,544 |
| 4 | × | × | × | | | × | | 161,909 | 0.723 | 21,365 |
| 5 | × | × | × | × | × | | | 877,366 | 0.077 | 7,936 |
| 6 | × | × | × | × | | × | | 181,214 | 0.614 | 3,210 |
| 7 | × | × | × | | × | × | | 152,553 | 0.852 | 14,078 |
| 8 | × | × | × | × | × | × | | 306,540 | 0.272 | 93 |
| **Fitted models with most favorable AIC** | | | | | | | | | | |
| Model | $X_1$ | $X_2$ | $X_3$ | $X_1X_2$ | $X_1X_3$ | $X_2X_3$ | $X_1X_2X_3$ | $\hat{N}$ | $\hat{\psi}$ | AIC |
| 9 | × | × | × | × | × | | × | 877,366 | 0.077 | 93 |
| 10 | × | × | × | × | | × | × | 181,214 | 0.614 | 93 |
| 11 | × | × | × | | × | × | × | 152,553 | 0.863 | 93 |
| 12 | × | × | | × | × | × | × | 203,973 | 0.500 | 93 |
| 13 | × | | × | × | × | × | × | 162,316 | 0.759 | 93 |
| 14 | | × | × | × | × | × | × | 154,311 | 0.842 | 93 |

*× indicates the predictor is included.

Table 3: Log-linear models for real three-stream data analyzed by Poorolajal, Mohammadi, and Farzinara (2017) under the usual conventions and with most favorable AIC

| **Fitted models under the usual conventions** | | | | | | | | | | |
|---|---|---|---|---|---|---|---|---|---|---|
| Model | $X_1$ | $X_2$ | $X_3$ | $X_1X_2$ | $X_1X_3$ | $X_2X_3$ | $X_1X_2X_3$ | $\widehat{N}$ | $\hat{\psi}$ | AIC |
| 1 | × | × | × | | | | | 10,421 | 0.040 | 112 |
| 2 | × | × | × | × | | | | 8,550 | 0.048 | 105 |
| 3 | × | × | × | | × | | | 9,751 | 0.044 | 111 |
| 4 | × | × | × | | | × | | 13,792 | 0.026 | 66 |
| 5 | × | × | × | × | × | | | 6,290 | 0.074 | 90 |
| 6 | × | × | × | × | | × | | 17,149 | 0.021 | 66 |
| 7 | × | × | × | | × | × | | 14,122 | 0.026 | 68 |
| 8 | × | × | × | × | × | × | | 41,922 | 0.008 | 60 |
| **Fitted models with most favorable AIC** | | | | | | | | | | |
| Model | $X_1$ | $X_2$ | $X_3$ | $X_1X_2$ | $X_1X_3$ | $X_2X_3$ | $X_1X_2X_3$ | $\widehat{N}$ | $\hat{\psi}$ | AIC |
| 9 | × | × | × | | | × | × | 14,904 | 0.024 | 59 |
| 10 | × | × | × | × | × | | × | 6,291 | 0.074 | 60 |
| 11 | × | × | × | × | | × | × | 17,149 | 0.021 | 60 |
| 12 | × | × | × | | × | × | × | 14,122 | 0.026 | 60 |
| 13 | × | × | | × | × | × | × | 2,600 | 0.500 | 60 |
| 14 | × | | × | × | × | × | × | 3,010 | 0.304 | 60 |
| 15 | | × | × | × | × | × | × | 3,353 | 0.229 | 60 |

*× indicates the predictor is included.

Table 4: Frequency of log-linear models selected by the AIC and averaged estimates from AIC-favored log-linear models across 1,000 simulations among 127 possible log-linear models and 8 possible log-linear models under the usual conventions

| **All possible 127 models** | | | | | | | | | |
|---|---|---|---|---|---|---|---|---|---|
| | $X_1$ | $X_2$ | $X_3$ | $X_1X_2$ | $X_1X_3$ | $X_2X_3$ | $X_1X_2X_3$ | Frequency [b] | Averaged Estimates of $N = 5{,}000$ |
| Scenario 1 [c] | × | × | | × | × | | × | 798 | 3,040 |
| | | × | × | × | | × | × | 52 | 3,804 |
| | × | × | × | × | × | × | | 150 | 8,052 |
| Scenario 2 [d] | × | × | × | × | × | | | 490 | 6,848 |
| | × | × | × | × | | | × | 1 | 6,232 |
| | × | × | | | × | × | × | 1 | 4,646 |
| | × | | × | | × | × | × | 453 | 4,407 |
| | × | × | × | × | × | × | | 55 | 6,871 |
| **8 possible models under the usual conventions** | | | | | | | | | |
| | $X_1$ | $X_2$ | $X_3$ | $X_1X_2$ | $X_1X_3$ | $X_2X_3$ | $X_1X_2X_3$ | Frequency | Averaged Estimates of $N = 5{,}000$ |
| Scenario 1 [c] | × | × | × | × | | × | | 4 | 4,089 |
| | × | × | × | × | × | × | | 996 | 8,053 |
| Scenario 2 [d] | × | × | × | × | × | | | 846 | 6,848 |
| | × | × | × | × | | × | | 1 | 4,409 |
| | × | × | × | × | × | × | | 153 | 6,871 |

*× indicates the predictor is included.

[b] The number of simulations that the model was selected by the AIC across 1,000 simulations.

[c] Data were generated from population-level multinomial model with $N = 5{,}000$, $p_1 = 0.3, p_{2|1} = 0.2, p_{2|\bar{1}} = 0.3, p_{3|12} = 0.8, p_{3|1\bar{2}} = 0.16, p_{3|\bar{1}2} = 0.5, \psi = 0.1$. Those parameters imply $E(N_{011}) = E(N_{010}) = 525$ and $\frac{p_{3|12}}{p_{3|\bar{1}2}} = \frac{p_{3|1\bar{2}}}{p_{3|\bar{1}\bar{2}}} = 1.6$.

[d] Data were generated from population-level multinomial model with $N = 5{,}000$, $p_1 = 0.3, p_{2|1} = 0.5, p_{2|\bar{1}} = 0.3, p_{3|12} = 0.35, p_{3|1\bar{2}} = 0.35, p_{3|\bar{1}2} = 0.25, \psi = 0.4375$. Those parameters imply $E(N_{111}) = E(N_{101}) = 262.5, E(N_{110}) = E(N_{100}) = 487.5, \frac{p_{3|1\bar{2}}}{p_{3|\bar{1}\bar{2}}} = 0.8$.

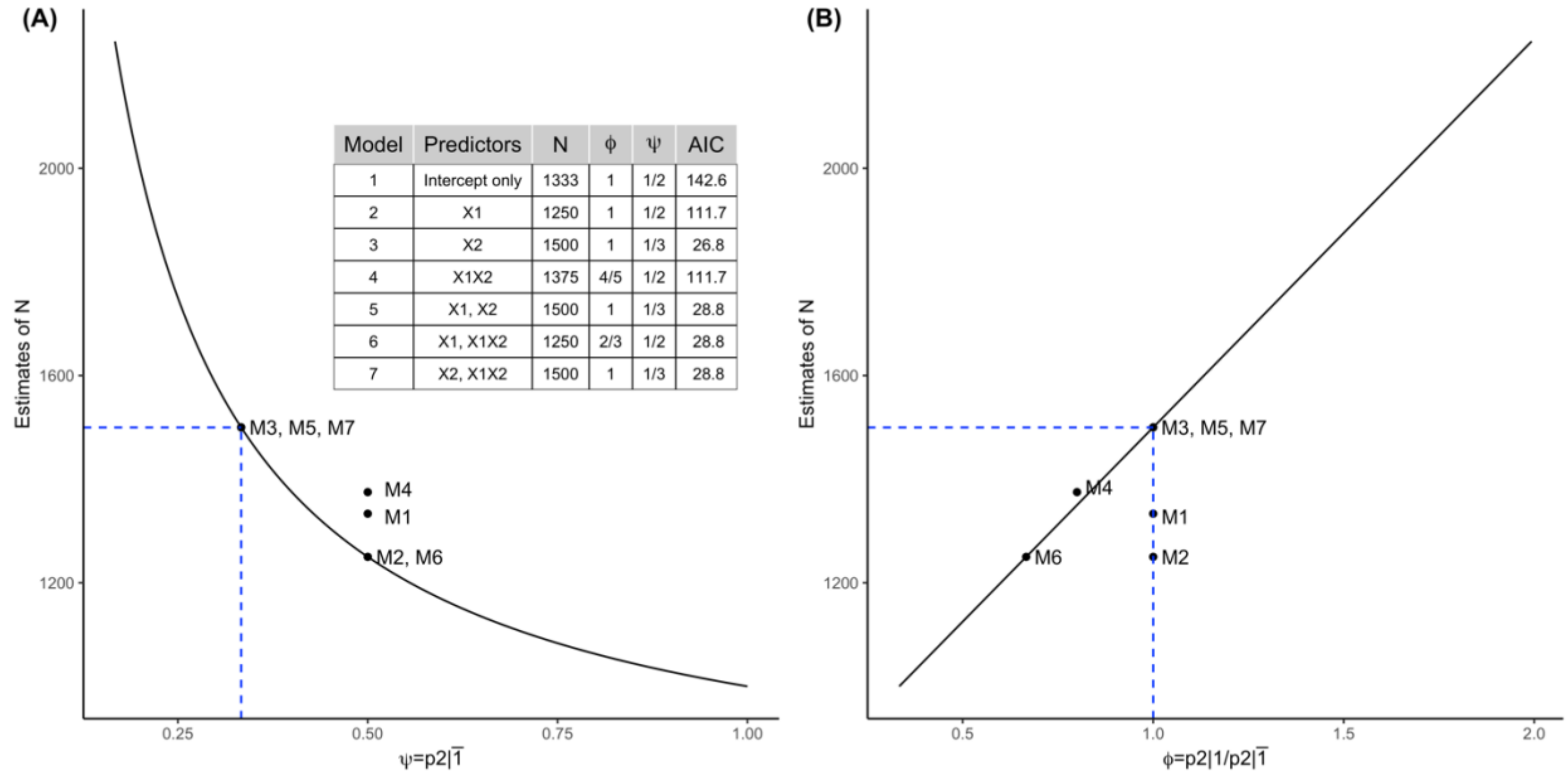

Figure 1: Estimates of $N$ from all possible log-linear models based on two-stream toy example CRC data presented in Table 1. Black solid line denotes the MLE of $N$ in Equation (1) with assumed $\psi = p_{2|\bar{1}}$ in panel (A); black solid line denotes the MLE of $N$ in Equation (2) in panel (B); black solid points represent estimates from all possible 7 log-linear models; blue dashed lines mark the estimates from the log-linear model with the lowest AIC (i.e., Model 3).

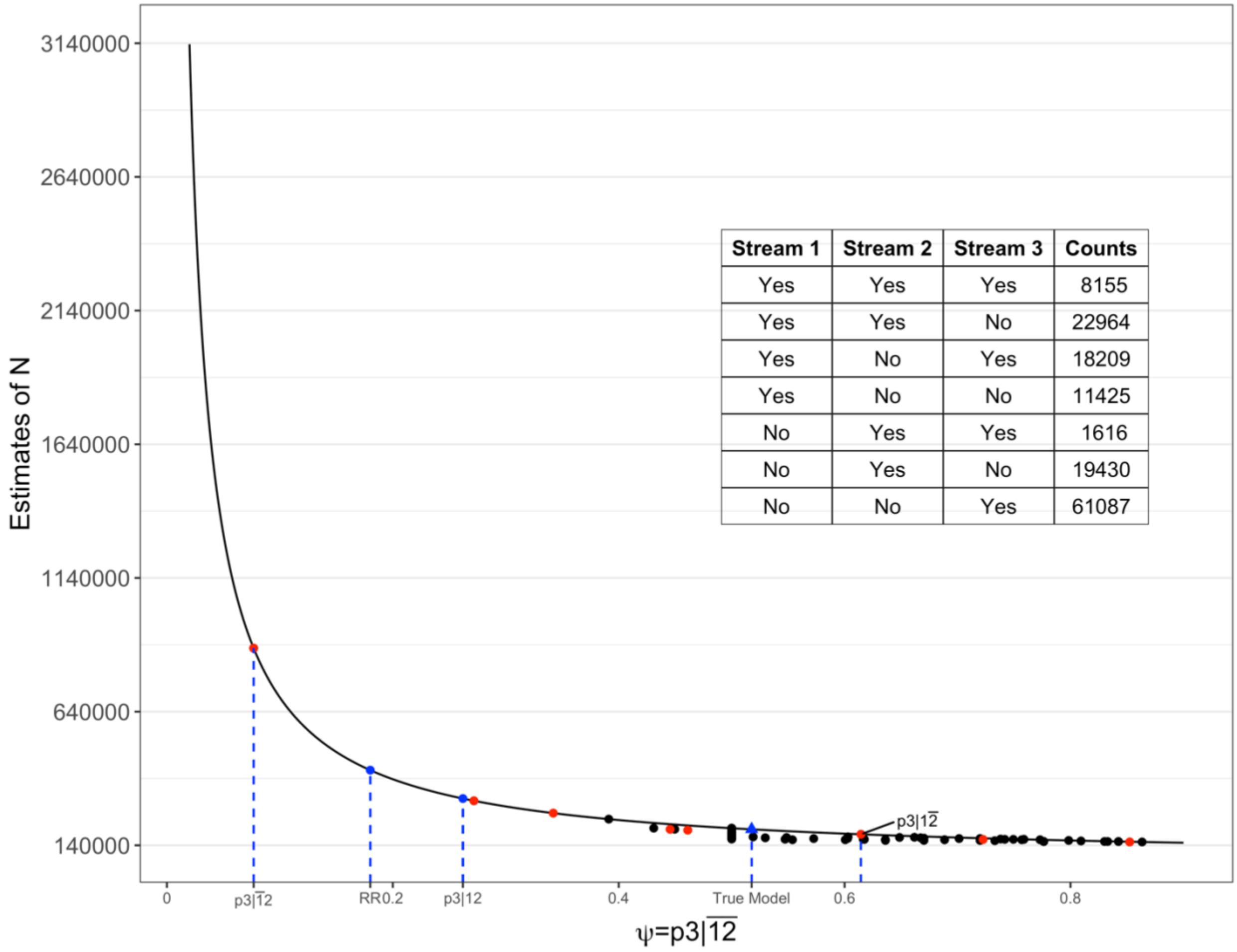

| Stream 1 | Stream 2 | Stream 3 | Counts |
|---|---|---|---|
| Yes | Yes | Yes | 8155 |
| Yes | Yes | No | 22964 |
| Yes | No | Yes | 18209 |
| Yes | No | No | 11425 |
| No | Yes | Yes | 1616 |
| No | Yes | No | 19430 |
| No | No | Yes | 61087 |

Figure 2: Estimates of $N$ from all possible log-linear models based on simulated data from the last column of Table 4 of Jones et al. (2014). Black solid line denotes the MLE of $N$ in Equation (3) as assumed $\psi = p_{3|\overline{12}}$ varies; red solid points denote estimates from the 8 possible log-linear models when imposing the usual conventions (i.e., no 3-way interaction and following the hierarchy principle); the union of red and black solid points denote estimates from all 127 possible log-linear models; blue solid points/dashed lines denote MLEs in Equation (3) under four different assumptions. On x-axis, RR denotes the assumption $\psi = \frac{p_{3|1\overline{2}} p_{3|\overline{1}2}}{p_{3|12}}$, $p_{3|12}$ denotes the assumption $\psi = p_{3|12}$. The text $p_{3|1\overline{2}}$ denotes the assumption $\psi = p_{3|1\overline{2}}$, and the text $p_{3|\overline{1}2}$ denotes the assumption $\psi = p_{3|\overline{1}2}$. The blue triangle/dashed line denotes the estimated $N$ by imposing correct assumptions, which assume S1 and S3 are independent conditional on S2 and 20% referral of individuals from S1 to S3.

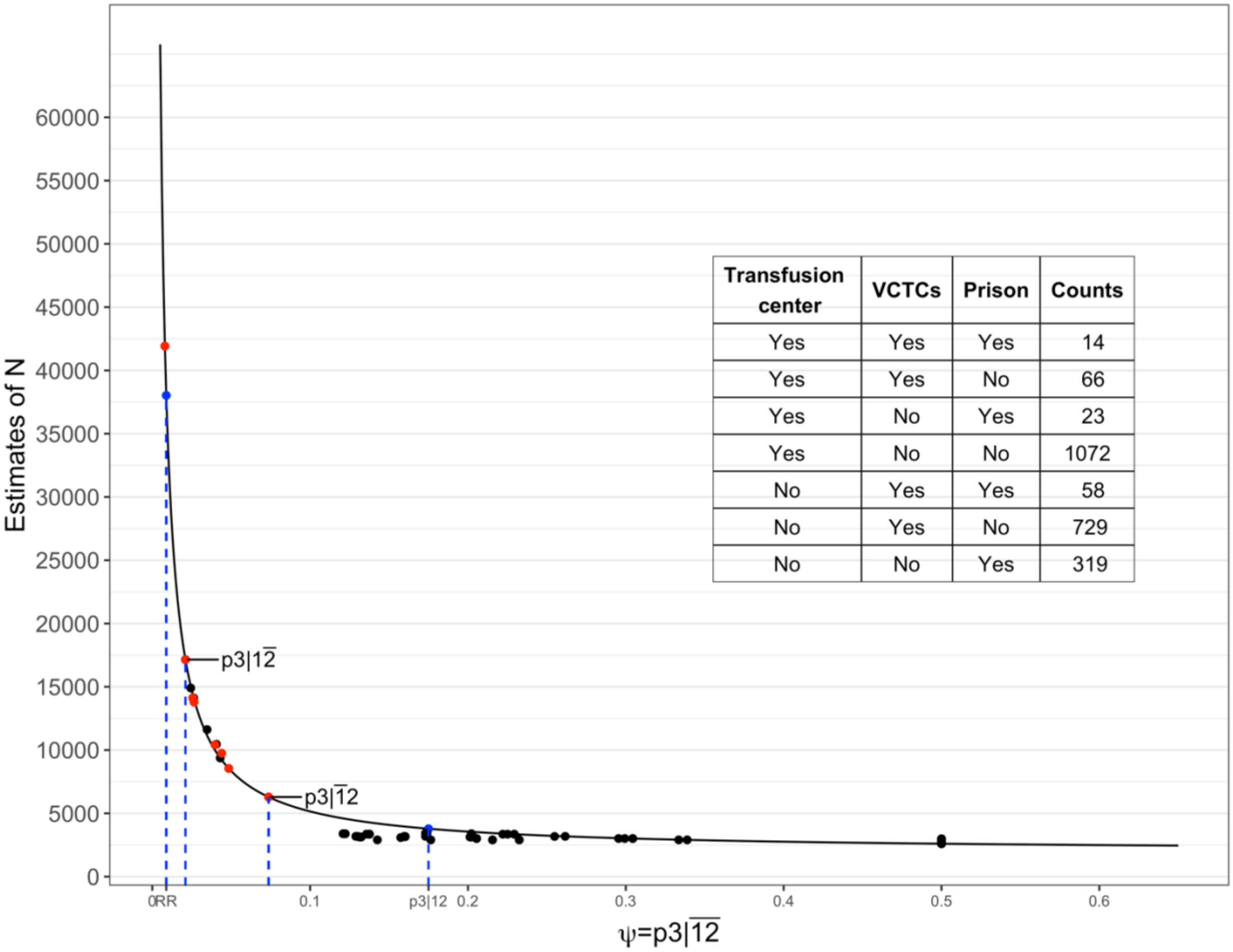

| Transfusion center | VCTCs | Prison | Counts |
|---|---|---|---|
| Yes | Yes | Yes | 14 |
| Yes | Yes | No | 66 |
| Yes | No | Yes | 23 |
| Yes | No | No | 1072 |
| No | Yes | Yes | 58 |
| No | Yes | No | 729 |
| No | No | Yes | 319 |

Figure 3: Estimates of the total HIV case count from all possible log-linear models based on data from Figure 1 of Poorolajal, Mohammadi, and Farzinara (2017), with "Transfusion center", "VCTCs", and "Prison" comprising Streams 1, 2, and 3, respectively. Black solid line denotes the MLE of $N$ in Equation (3) as assumed $\psi = p_{3|\overline{12}}$ varies; red solid points denote estimates from the 8 possible log-linear models when imposing the usual conventions (i.e., no 3-way interaction and following the hierarchy principle); the union of red and black solid points denote estimates from all 127 possible log-linear models; blue solid points/dashed lines denote MLEs in Equation (1) under four different assumptions. On x-axis, RR denotes the assumption $\psi = \frac{p_{3|1\bar{2}} p_{3|\bar{1}2}}{p_{3|12}}$, $p_{3|12}$ denotes the assumption $\psi = p_{3|12}$. The text $p_{3|1\bar{2}}$ denotes the assumption $\psi = p_{3|1\bar{2}}$, and the text $p_{3|\bar{1}2}$ denotes the assumption $\psi = p_{3|\bar{1}2}$.

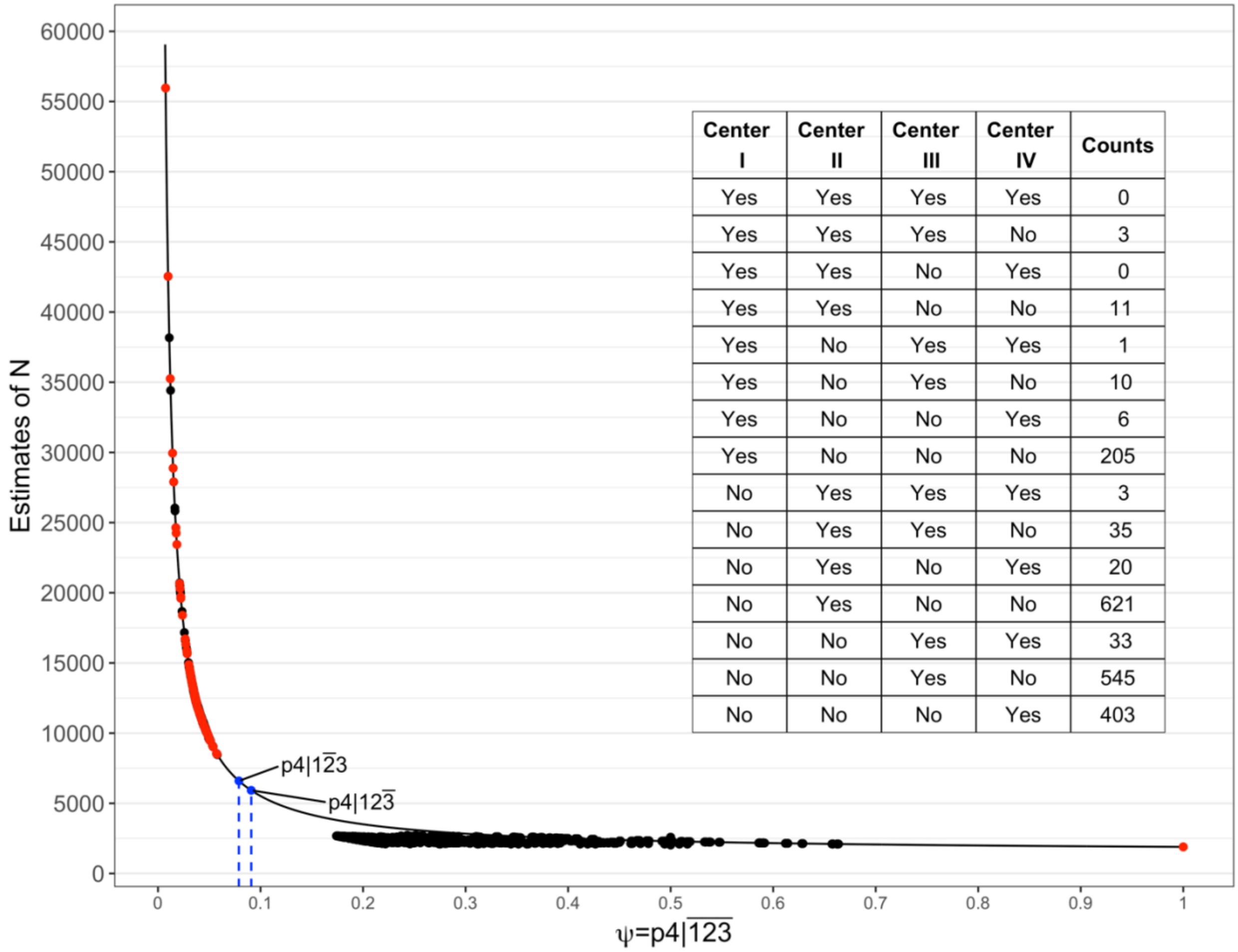

| Center I | Center II | Center III | Center IV | Counts |
|---|---|---|---|---|
| Yes | Yes | Yes | Yes | 0 |
| Yes | Yes | Yes | No | 3 |
| Yes | Yes | No | Yes | 0 |
| Yes | Yes | No | No | 11 |
| Yes | No | Yes | Yes | 1 |
| Yes | No | Yes | No | 10 |
| Yes | No | No | Yes | 6 |
| Yes | No | No | No | 205 |
| No | Yes | Yes | Yes | 3 |
| No | Yes | Yes | No | 35 |
| No | Yes | No | Yes | 20 |
| No | Yes | No | No | 621 |
| No | No | Yes | Yes | 33 |
| No | No | Yes | No | 545 |
| No | No | No | Yes | 403 |

Figure 4: Estimates of the total number HIV cases from all possible log-linear models based on four-stream CRC data from Table 2 of Abeni, Brancato, and Perucci (1994), with "Center I", "Center II" and "Center III", and "Center IV" comprising Streams 1, 2, 3, and 4, respectively. Black solid line denotes the MLE of $N$ in Equation (3) as assumed $\psi = p_{4|\bar{1}\bar{2}\bar{3}}$ varies; red solid points denote estimates from the 113 possible log-linear models when imposing the usual conventions (i.e., no 4-way interaction and following the hierarchy principle); the union of red and black solid points denote estimates from all possible 32,767 log-linear models. The text $p_{4|1\bar{2}3}$ denotes the assumption $\psi = p_{4|1\bar{2}3}$, and the text $p_{4|12\bar{3}}$ denotes the assumption $\psi = p_{4|12\bar{3}}$.

# Supplementary Materials for "On some pitfalls of the log-linear modeling framework for capture-recapture studies in disease surveillance"

Yuzi Zhang[1,*], Lin Ge[1], Lance A. Waller[1], and Robert H. Lyles[1]
[1]Department of Biostatistics and Bioinformatics, The Rollins School of Public Health of Emory University, 1518 Clifton Rd. N.E., Atlanta, GA, USA
*yuzi.zhang@emory.edu

### Simulation Settings

Data were generated from the population-level multinomial:
$(N_{111}, N_{110}, N_{101}, N_{100}, N_{011}, N_{010}, N_{001}, N_{000}) \sim \text{Multinominal}(N, p_{111}, p_{110}, p_{101}, p_{100}, p_{011}, p_{010}, p_{001}, p_{000})$.
The true number of cases $N$ is set to $5{,}000$ under both scenarios, and capture probabilities are computed based on parameters $(p_1, p_{2|1}, p_{2|\bar{1}}, p_{3|12}, p_{3|1\bar{2}}, p_{3|\bar{1}2}, p_{3|\overline{12}})$:

$$p_{111} = p_1 p_{2|1} p_{3|12}$$
$$p_{110} = p_1 p_{2|1}(1 - p_{3|12})$$
$$p_{101} = p_1(1 - p_{2|1}) p_{3|1\bar{2}}$$
$$p_{100} = p_1(1 - p_{2|1})(1 - p_{3|1\bar{2}})$$
$$p_{011} = (1 - p_1) p_{2|\bar{1}} p_{3|\bar{1}2}$$
$$p_{010} = (1 - p_1) p_{2|\bar{1}}(1 - p_{3|\bar{1}2})$$
$$p_{000} = (1 - p_1)(1 - p_{2|\bar{1}})(1 - p_{3|\overline{12}}),$$

where under three-stream cases, $\psi = p_{3|\overline{12}}$.

*Scenario 1*

We assume the probability of having capture history $(0,1,1)$ is equal to the probability of having capture history $(0,1,0)$, and that the association between the first data stream and the third data stream is not affected by whether cases are identified by the second data stream. Converting to mathematical expressions, these stipulations correspond to setting the testable assumption $p_{011} = p_{010}$ i.e., $E(N_{011}) = E(N_{010})$, and the untestable assumption $p_{3|12}/p_{3|\bar{1}2} = p_{3|1\bar{2}}/\psi$. True values of the parameters are: $p_1 = 0.3, p_{2|1} = 0.2, p_{2|\bar{1}} = 0.3, p_{3|12} = 0.8, p_{3|1\bar{2}} = 0.16, p_{3|\bar{1}2} = 0.5, \psi = 0.1$.

*Scenario 2*

We impose two testable assumptions $E(N_{111}) = E(N_{101})$ and $E(N_{110}) = E(N_{100})$, and one untestable assumption which states that the key parameter $\psi = p_{3|1\bar{2}}/0.8$. This untestable assumption implies that, among those not identified by the second stream, cases are more likely to be captured by the third stream if they are not captured by the first stream, i.e., the first stream and third stream are negatively correlated conditional on a lack of capture by the second stream.

Table S1: Possible log-linear models for two-stream toy example data in Table 1

| | | Fitted cell counts [b] | | | | MLE of key parameters [c] | | Results from the toy example data presented in Table 1 [d] | | | |
|---|---|---|---|---|---|---|---|---|---|---|---|
| Model [a] | Predictors | $\hat{N}_{11}$ | $\hat{N}_{10}$ | $\hat{N}_{01}$ | $\hat{N}_{00}$ | $\hat{\psi}$ | $\hat{\phi}$ | $\hat{\psi}$ | $\hat{\phi}$ | $\hat{N}$ | AIC |
| 1 | Intercept only | $\frac{n_c}{3}$ | $\frac{n_c}{3}$ | $\frac{n_c}{3}$ | $\frac{n_c}{3}$ | $\frac{1}{2}$ | 1 | $\frac{1}{2}$ | 1 | 1333 | 142.6 |
| 2 | $X_1$ | $\frac{n_{1\cdot}}{2}$ | $\frac{n_{1\cdot}}{2}$ | $n_{01}$ | $n_{01}$ | $\frac{1}{2}$ | 1 | $\frac{1}{2}$ | 1 | 1250 | 111.7 |
| 3 | $X_2$ | $\frac{n_{\cdot 1}}{2}$ | $n_{10}$ | $\frac{n_{\cdot 1}}{2}$ | $n_{10}$ | $\frac{n_{\cdot 1}}{2n_{10}+n_{\cdot 1}}$ | 1 | $\frac{1}{3}$ | 1 | 1500 | 26.8 |
| 4 | $X_1X_2$ | $n_{11}$ | $\frac{n_{10}+n_{01}}{2}$ | $\frac{n_{10}+n_{01}}{2}$ | $\frac{n_{10}+n_{01}}{2}$ | $\frac{1}{2}$ | $\frac{4n_{11}}{2n_{11}+n_{10}+n_{01}}$ | $\frac{1}{2}$ | 0.8 | 1375 | 111.7 |
| 5 | $X_1, X_2$ | $n_{11}$ | $n_{10}$ | $n_{01}$ | $\frac{n_{10}n_{01}}{n_{11}}$ | $\frac{n_{11}}{n_{11}+n_{10}}$ | 1 | $\frac{1}{3}$ | 1 | 1500 | 28.8 |
| 6 | $X_1, X_1X_2$ | $n_{11}$ | $n_{10}$ | $n_{01}$ | $n_{01}$ | $\frac{1}{2}$ | $\frac{2n_{11}}{n_{1\cdot}}$ | $\frac{1}{2}$ | $\frac{2}{3}$ | 1250 | 28.8 |
| 7 | $X_2, X_1X_2$ | $n_{11}$ | $n_{10}$ | $n_{01}$ | $n_{10}$ | $\frac{n_{01}}{n_{01}+n_{10}}$ | $\frac{n_{11}(n_{10}+n_{01})}{n_{01}n_{1\cdot}}$ | $\frac{1}{3}$ | 1 | 1500 | 28.8 |

[a] The intercept ($\alpha$) is included in all models

[b] $n_c = n_{11} + n_{10} + n_{01}$, $n_{1\cdot} = n_{11} + n_{10}$, $n_{\cdot 1} = n_{11} + n_{01}$; analytic results in columns 3 through 8 reproduced from Lyles et al. (2021)

[c] $\hat{\psi} = \frac{\hat{N}_{10}}{\hat{N}_{10}+\hat{N}_{00}}$ and $\hat{\phi} = \frac{\hat{N}_{11}(\hat{N}_{01}+\hat{N}_{00})}{(\hat{N}_{11}+\hat{N}_{10})\hat{N}_{01}}$

[d] $\hat{N} = n_c + \exp(\hat{\alpha})$, where $\hat{\alpha}$ is the estimated intercept from fitting the log-linear model based on the toy example data presented in Table 1

Table S2: Possible log-linear models for three-stream CRC data when applying the usual conventions

| Model [a] | Predictors | MLE of key parameter $\psi = p_{3\|\overline{12}}$ [b] | Fitted $N_{000}$ [c] |
|---|---|---|---|
| 1 | $X_1, X_2, X_3$ | $\frac{\hat{N}_{111}^{\frac{3}{8}}\hat{N}_{101}^{\frac{1}{4}}\hat{N}_{011}^{\frac{1}{4}}\hat{N}_{001}^{\frac{1}{8}}}{\hat{N}_{111}^{\frac{3}{8}}\hat{N}_{101}^{\frac{1}{4}}\hat{N}_{011}^{\frac{1}{4}}\hat{N}_{001}^{\frac{1}{8}} + \hat{N}_{110}^{\frac{1}{4}}\hat{N}_{100}^{\frac{3}{8}}\hat{N}_{010}^{\frac{3}{8}}}$ | $\frac{\hat{N}_{100}^{\frac{1}{2}}\hat{N}_{010}^{\frac{1}{2}}\hat{N}_{001}^{\frac{1}{2}}}{\hat{N}_{111}^{\frac{1}{2}}}$ |
| 2 | $X_1, X_2, X_3, X_1X_2$ | $\frac{\hat{N}_{111}^{\frac{1}{3}}\hat{N}_{101}^{\frac{1}{3}}\hat{N}_{011}^{\frac{1}{3}}}{\hat{N}_{111}^{\frac{1}{3}}\hat{N}_{101}^{\frac{1}{3}}\hat{N}_{011}^{\frac{1}{3}} + \hat{N}_{110}^{\frac{1}{3}}\hat{N}_{100}^{\frac{1}{3}}\hat{N}_{010}^{\frac{1}{3}}}$ | $\frac{\hat{N}_{110}^{\frac{1}{3}}\hat{N}_{100}^{\frac{1}{3}}\hat{N}_{010}^{\frac{1}{3}}}{\hat{N}_{111}^{\frac{1}{3}}\hat{N}_{101}^{\frac{1}{3}}\hat{N}_{011}^{\frac{1}{3}}}\hat{N}_{001}$ |
| 3 | $X_1, X_2, X_3, X_1X_3$ | $\frac{\hat{N}_{111}^{\frac{1}{6}}\hat{N}_{110}^{\frac{1}{6}}\hat{N}_{011}^{\frac{2}{3}}\hat{N}_{001}^{\frac{1}{3}}}{\hat{N}_{111}^{\frac{1}{6}}\hat{N}_{110}^{\frac{1}{6}}\hat{N}_{011}^{\frac{2}{3}}\hat{N}_{001}^{\frac{1}{3}} + \hat{N}_{101}^{\frac{1}{6}}\hat{N}_{100}^{\frac{1}{6}}\hat{N}_{010}}$ | $\frac{\hat{N}_{101}^{\frac{1}{3}}\hat{N}_{100}^{\frac{1}{3}}\hat{N}_{001}^{\frac{1}{3}}}{\hat{N}_{111}^{\frac{1}{3}}\hat{N}_{110}^{\frac{1}{3}}\hat{N}_{011}^{\frac{1}{3}}}\hat{N}_{010}$ |
| 4 | $X_1, X_2, X_3, X_2X_3$ | $\frac{\hat{N}_{111}^{\frac{1}{6}}\hat{N}_{110}^{\frac{1}{6}}\hat{N}_{101}^{\frac{2}{3}}\hat{N}_{001}^{\frac{1}{3}}}{\hat{N}_{111}^{\frac{1}{6}}\hat{N}_{110}^{\frac{1}{6}}\hat{N}_{101}^{\frac{2}{3}}\hat{N}_{001}^{\frac{1}{3}} + \hat{N}_{100}\hat{N}_{011}^{\frac{1}{6}}\hat{N}_{010}^{\frac{1}{6}}}$ | $\frac{\hat{N}_{011}^{\frac{1}{3}}\hat{N}_{010}^{\frac{1}{3}}\hat{N}_{001}^{\frac{1}{3}}}{\hat{N}_{111}^{\frac{1}{3}}\hat{N}_{110}^{\frac{1}{3}}\hat{N}_{101}^{\frac{1}{3}}}\hat{N}_{100}$ |
| 5 | $X_1, X_2, X_3, X_1X_2, X_1X_3$ | $\frac{\hat{N}_{011}}{\hat{N}_{011} + \hat{N}_{010}}$ | $\frac{\hat{N}_{010}\hat{N}_{001}}{\hat{N}_{011}}$ |
| 6 | $X_1, X_2, X_3, X_1X_2, X_2X_3$ | $\frac{\hat{N}_{101}}{\hat{N}_{101} + \hat{N}_{100}}$ | $\frac{\hat{N}_{100}\hat{N}_{001}}{\hat{N}_{101}}$ |
| 7 | $X_1, X_2, X_3, X_1X_3, X_2X_3$ | $\frac{\hat{N}_{110}\hat{N}_{101}^{\frac{1}{4}}\hat{N}_{011}^{\frac{1}{4}}\hat{N}_{001}^{\frac{3}{4}}}{\hat{N}_{110}\hat{N}_{101}^{\frac{1}{4}}\hat{N}_{011}^{\frac{1}{4}}\hat{N}_{001}^{\frac{3}{4}} + \hat{N}_{111}^{\frac{1}{4}}\hat{N}_{100}\hat{N}_{010}}$ | $\frac{\hat{N}_{100}\hat{N}_{010}}{\hat{N}_{110}}$ |
| 8 | $X_1, X_2, X_3, X_1X_2, X_1X_3, X_2X_3$ | $\frac{\hat{N}_{110}\hat{N}_{101}\hat{N}_{011}}{\hat{N}_{110}\hat{N}_{101}\hat{N}_{011} + \hat{N}_{111}\hat{N}_{100}\hat{N}_{011}}$ | $\frac{\hat{N}_{111}\hat{N}_{100}\hat{N}_{010}\hat{N}_{001}}{\hat{N}_{110}\hat{N}_{101}\hat{N}_{011}}$ |

[a] The intercept ($\alpha$) is included in all models

[b] $\hat{N}_{ijk}$ denotes the fitted cell count with capture history $(i, j, k)$ and is obtained by computing estimated $E(N_{ijk})$ from the fitted log-linear model, where $i, j, k \in \{0,1\}$; Under models 1- 4, and 7 (which are unsaturated models), fitted cell counts do not have closed form and can be computed by numerically maximize the Poisson log-likelihood; while fitted cell counts $\hat{N}_{011} = n_{011}$ and $\hat{N}_{010} = n_{010}$ under the model 5, and fitted cell counts $\hat{N}_{101} = n_{101}$ and $\hat{N}_{100} = n_{100}$ under the model 6; the saturated model 8 yields each of fitted cell counts equal to its corresponding observed cell count.

[c] Fitted $N_{000}$ is computed as $\exp(\hat{\alpha})$, where $\hat{\alpha}$ is the MLE of $\alpha$